\documentclass{PoS}

\def\MET{$/$\hspace{-1.3ex}$E_T$}

\PoS{PoS(HEP2005)348}

\title{Searches for Scalar Top and Bottom Quarks at the Tevatron}

\ShortTitle{Searches for Scalar Top and Bottom Quarks at the Tevatron}

\author{\speaker{Thomas Nunnemann}\thanks{for the D\O{} and CDF Collaborations}\\
        LMU Munich, Germany\\
        E-mail: \email{Thomas.Nunnemann@physik.uni-muenchen.de}}

\abstract{Searches for the supersymmetric partners of top and bottom quarks
  using data up to 
  340\,pb$^{-1}$ taken at the Tevatron $p\bar{p}$ collider are described.
  We report on searches for scalar top quarks $\tilde{t}$ in the decays
  $\tilde{t}\rightarrow c \tilde{\chi}_1^0$ and 
  $\tilde{t}\rightarrow bl\tilde{\nu}$
  and for scalar bottom quarks $\tilde{b}$ in the decay 
  $\tilde{b}\rightarrow b \tilde{\chi}_1^0$. No evidence for a signal has been
  found, but improved exclusion regions have been derived in the 
  framework of a generic minimal superymmetric extension of the
  standard model.}

\FullConference{International Europhysics Conference on High Energy Physics\\
		 July 21st - 27th 2005\\
		 Lisboa, Portugal}

\begin{document}

\section{Introduction}
A major consequence of the realization of supersymmetry (SUSY) in nature
would be the existence of scalar partner particles of the standard model
fermions.
The pair-production of scalar quarks (squarks) at the Tevatron can proceed
through
gluon fusion or quark annihilation and could have a significant cross-section
for relatively low squark masses.

Large Higgs Yukawa couplings to the third quark generation induce a strong 
mixing between the supersymmetric partners of the two chirality states
of the top (and bottom) quark, which leads to two physical states,
$\tilde{t}_1$ and  $\tilde{t}_2$ (and $\tilde{b}_1$, $\tilde{b}_2$),
of different mass~\cite{Ellis}. Therefore the lightest scalar top quark
$\tilde{t}_1$ could possibly be much lighter than other squarks.
At large values of $\tan\beta$ a relatively light $\tilde{b}_1$ is also
expected.

If kinematically allowed, a squark will predominantly decay via 
$\tilde{q}\rightarrow q \tilde{\chi}_1^0$, leading to a topology
of two jets and missing transverse energy~\cite{Munar}. In the case of
the sbottom quark decay the background can be efficiently suppressed by
requiring a $b$-tag. The two-body stop decays 
$\tilde{t}_1 \rightarrow t \tilde{\chi}^0_1$ and
$\tilde{t}_1 \rightarrow b \tilde{\chi}^+_1$ are kinematically forbidden
in the accessible parameter region at the Tevatron. Here, the most promising
searches are based on the loop-induced decay 
$\tilde{t}_1 \rightarrow c \tilde{\chi}^0_1$
and the three-body decay $\tilde{t}_1 \rightarrow b l \tilde{\nu}$. The latter
might be favored over the decay $\tilde{t}_1 \rightarrow b W \tilde{\chi}_1^0$
due to the relatively weak constraint on the sneutrino mass 
$M(\tilde{\nu})>43.7$\,GeV~\cite{lepsusy}.

\section{Search for Scalar Top Quarks in the Decay 
$\tilde{t}_1 \rightarrow c \tilde{\chi}^0_1$}
The CDF and D\O{} Collaborations have searched for scalar top quarks decaying
into $c \tilde{\chi}^0_1$ in approximately 90\,pb$^{-1}$ of Run I
data~\cite{CDF1,D01}. The CDF collaboration also performed a preliminary
measurement based on 163\,pb$^{-1}$ of Run II data.

\begin{figure}[b]
\includegraphics[width=.39\textwidth]{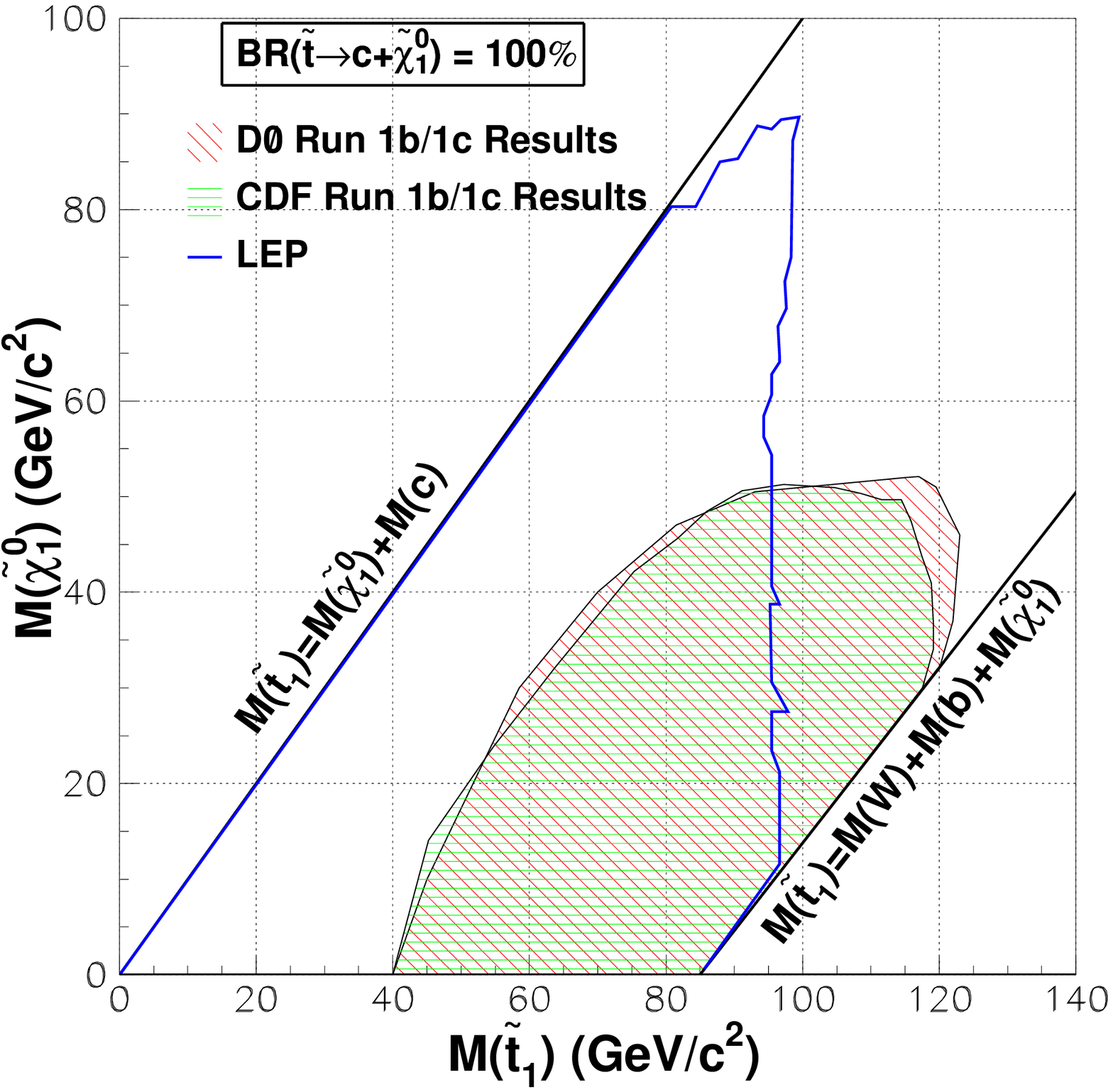}
\hspace{0.03\textwidth}
\includegraphics[width=.58\textwidth]{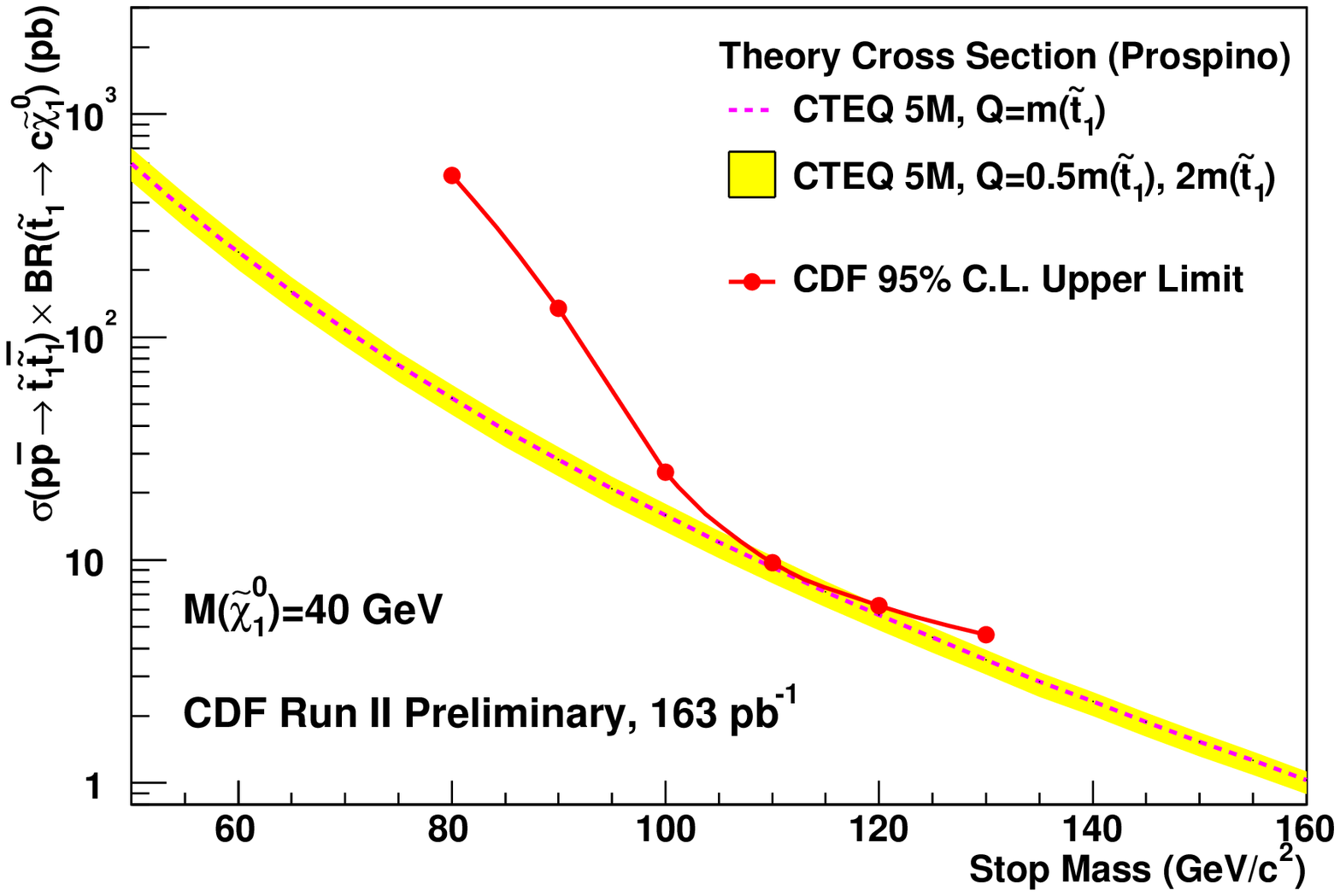}
\caption{Exclusion limits for $\tilde{t}_1 \rightarrow c \tilde{\chi}^0_1$.
{\it Left:} 95\% C.L. exclusion region as function of $M_{\tilde{t}}$ and
$M_{\tilde{\chi}_1^0}$. {\it Right:} Cross-section limit $\tilde{t}\bar{\tilde{t}}$
pair production for assumed $M_{\tilde{\chi}_1^0}=40$\,GeV compared to NLO
prediction.} 
\label{fig1}
\end{figure}

Candidate events are selected requiring significant missing transverse
energy \MET{} and two reconstructed jets. The D\O{} analysis does
not attempt to identify the flavor of the jets, whereas the CDF measurements employ
a heavy-flavor tag based on the probability that all the tracks in the jet
come from the primary vertex. The background from $W$ production in association
with jets is suppressed by vetoing events with an isolated lepton. 
Other backgrounds are $Z$ production in association with jets (with
$Z\rightarrow \nu\bar{\nu}$) and QCD multi-jet production. No excess over
standard model background has been observed and exclusion limits 
as a function of $M_{\tilde{t}}$ and $M_{\tilde{\chi}_1^0}$ have been
generically derived within the minimal supersymmetric extension of the standard
model (MSSM)~\cite{Martin}. 

The limits are shown in Fig.~\ref{fig1}, left,
together with the combined LEP result~\cite{lepsusy}. The sensitivity of the
Tevatron measurements decreases close to the kinematic limit
$M_{\tilde{t}}=M_{\tilde{\chi}_1^0}+M_c$ due to the minimal required 
transverse energy of
the jets ($E_T$) and \MET{}. CDF's preliminary Run II upper cross-section limits 
shown in Fig.~\ref{fig1}, right, do not provide additional constraints on the
stop mass, but projections based on integrated luminosities up to 
4\,fb$^{-1}$ indicate sensitivities up to $M_{\tilde{t}}\approx 175$\,GeV
at $M_{\tilde{\chi}_1^0}\approx 100$\,GeV.

\section{Search for Scalar Top Quarks in the Decay 
$\tilde{t}_1 \rightarrow b l \tilde{\nu}$}
The best sensitivity for stop pair-production with subsequent decays
$\tilde{t}_1 \rightarrow b l \tilde{\nu}$
is obtained in the $e\mu$ channel. Its branching ratio is twice as large as
for the $ee$ or $\mu\mu$ channels and the background from $Z$ and
Drell-Yan production is largely reduced.

In Run II, the only preliminary result so far has been obtained in the
less preferred $\mu\mu$ channel using 340\,pb$^{-1}$ of data, taken with the
D\O{} experiment. To maximize the sensitivity close to the kinematic boundary,
muons with transverse momenta as low as $p_T(\mu_1)>8$\,GeV and 
$p_T(\mu_2)>6$\,GeV for the leading and trailing muon, respectively, are
accepted. In addition one jet with $b$-tag and $E_T>15$\,GeV is required. 
After applying a $Z$-veto and a two-dimensional cut on \MET{} and the angle 
between \MET{} and the leading muon, 
the background is dominated by top pair-production.
The systematic error is dominated by uncertainties in the jet energy scale
and the $b$-tagging efficiency.

\begin{figure}
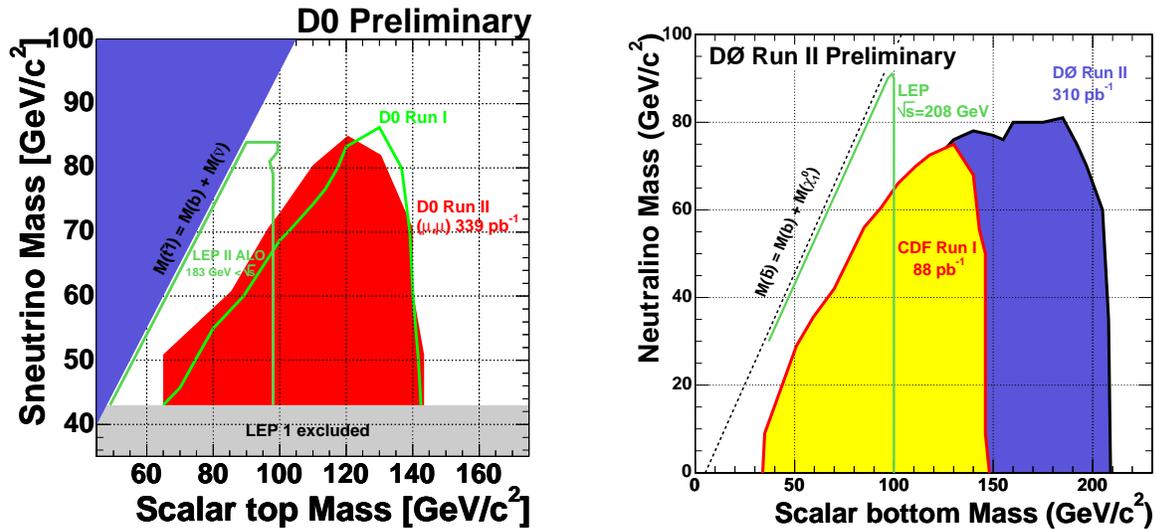

\includegraphics[width=.46\textwidth]{stop_blsnu_excl.epsi}
\hspace{0.08\textwidth}\includegraphics[width=.46\textwidth]{sbottom_excl.epsi}
\caption{95\% C.L. exclusion regions for the 
$\tilde{t}_1 \rightarrow b l \tilde{\nu}$ (left) and 
$\tilde{b}\rightarrow b \tilde{\chi}_1^0$ search (right).}
\label{fig2}
\end{figure}

The shape of the scalar sum of the jet transverse energies
$H_T=\sum_\mathrm{jets} E_T$ is used to further discriminate between the
$\tilde{t}\bar{\tilde{t}}$ signal and the standard model $t\bar{t}$ 
background. 
Cross-section limits are calculated with the likelihood ratio 
method~\cite{Junk} assuming a branching ratio 
$BR(\tilde{t} \rightarrow b l \tilde{\nu})=100\%$.
By comparing them to next-to-leading order predictions of the signal 
cross-section calculated using Prospino 2~\cite{Beenakker}, 
exclusion regions in
the plane given by $M_{\tilde{t}}$ and $M_{\tilde{\nu}}$ are derived
(cf. Fig.~\ref{fig2}, left). 
Compared to previous measurments
an additional region at low 
$\Delta M(\tilde{t},\tilde{\nu})$ has been excluded due to the low
muon $p_T$ requirements. 
A significant extension of the exclusion limit
is expected from a pending analysis using the preferred $e\mu$
channel.

\section{Search for Scalar Bottom Quarks in the Decay
$\tilde{b}\rightarrow b \tilde{\chi}_1^0$}
D\O{} has searched for sbottom pair-production in 310\,pb$^{-1}$
of data taken with a dedicated trigger designed for topologies with both
significant \MET{} and jet activity, which is based
on the vector sum of the jet momenta and acoplanarity.

The signal selection requires \MET{} and two or three reconstructed jets with
one tight $b$-tag based on the jet lifetime probability. Events with more
than three jets are rejected to minimize background from top pair-production
and a veto on isolated $e$, $\mu$, and tracks (originating from $\tau$ decays)
is applied to suppress vector boson production. Background from QCD 
multi-jet production has been shown to vanish for large \MET{}.
The average \MET{} and jet $E_T$ expected for the signal largely depends on
the mass difference of sbottom quark and neutralino. Therefore three 
sets of cuts on these
variables have been optimized for different assumed $M_{\tilde{b}}$.
%The leading (second) 
%jet is required to have a minimal $E_T$ ranging from 40-70\,GeV (15-40\,GeV)
%and missi

As with the stop searches, the result is interpreted in the
framework of generic MSSM and a 95\% exclusion region in the 
$(M_{\tilde{b}}, M_{\tilde{\chi}_1^0})$-plane is derived, shown 
in Fig.~\ref{fig2}, right. The excluded region is significantly increased
compared to previous measurements~\cite{lepsusy,CDF1}, corresponding to a gain
of approximately 60\,GeV in the sbottom mass limit at fixed
$M_{\tilde{\chi}_1^0}$.

\section{Conclusions}
Searches for scalar top and bottom quarks at the Tevatron have now a 
significantly increased sensitivity compared to previous results obtained
at LEP and Run I. This has been achieved not only because of the increased
Tevatron luminosity and beam energy, but also due to strengthened
heavy-flavor tagging and the ability to loosen the requirement on the
lepton momenta. Further improvements are expected from an anticipated
reduction of the jet energy scale uncertainty, the inclusion of additional
decay channels, and with the increasing data set of Run II.

\end{document}